\RequirePackage{lineno}
\documentclass[twocolumn,showpacs,preprintnumbers,superscriptaddress,amsmath,amssymb]{revtex4}
\newcommand{\pT}{\ensuremath{p_{T}}}
\newcommand{\mT}{\ensuremath{m_{T}}}
\newcommand{\pp}{\ensuremath{pp}}
\newcommand{\snn}{\ensuremath{\sqrt{s_{NN}}}}
\newcommand{\npart}{\ensuremath{N_{\rm part}}}
\newcommand{\nch}{\ensuremath{dN_{\rm ch}/d\eta}}
\newcommand{\nchy}{\ensuremath{dN_{\rm ch}/dy}}

\usepackage{graphicx}
\usepackage{dcolumn}
\usepackage{bm}
\usepackage{color}

\begin{document}

\preprint{Version 15}

\title{Scaling properties at freeze-out in relativistic heavy ion collisions}

\date{\today}

\affiliation{Argonne National Laboratory, Argonne, Illinois 60439, USA}
\affiliation{University of Birmingham, Birmingham, United Kingdom}
\affiliation{Brookhaven National Laboratory, Upton, New York 11973, USA}
\affiliation{University of California, Berkeley, California 94720, USA}
\affiliation{University of California, Davis, California 95616, USA}
\affiliation{University of California, Los Angeles, California 90095, USA}
\affiliation{Universidade Estadual de Campinas, Sao Paulo, Brazil}
\affiliation{University of Illinois at Chicago, Chicago, Illinois 60607, USA}
\affiliation{Creighton University, Omaha, Nebraska 68178, USA}
\affiliation{Czech Technical University in Prague, FNSPE, Prague, 115 19, Czech Republic}
\affiliation{Nuclear Physics Institute AS CR, 250 68 \v{R}e\v{z}/Prague, Czech Republic}
\affiliation{University of Frankfurt, Frankfurt, Germany}
\affiliation{Institute of Physics, Bhubaneswar 751005, India}
\affiliation{Indian Institute of Technology, Mumbai, India}
\affiliation{Indiana University, Bloomington, Indiana 47408, USA}
\affiliation{Alikhanov Institute for Theoretical and Experimental Physics, Moscow, Russia}
\affiliation{University of Jammu, Jammu 180001, India}
\affiliation{Joint Institute for Nuclear Research, Dubna, 141 980, Russia}
\affiliation{Kent State University, Kent, Ohio 44242, USA}
\affiliation{University of Kentucky, Lexington, Kentucky, 40506-0055, USA}
\affiliation{Institute of Modern Physics, Lanzhou, China}
\affiliation{Lawrence Berkeley National Laboratory, Berkeley, California 94720, USA}
\affiliation{Massachusetts Institute of Technology, Cambridge, MA 02139-4307, USA}
\affiliation{Max-Planck-Institut f\"ur Physik, Munich, Germany}
\affiliation{Michigan State University, East Lansing, Michigan 48824, USA}
\affiliation{Moscow Engineering Physics Institute, Moscow Russia}
\affiliation{NIKHEF and Utrecht University, Amsterdam, The Netherlands}
\affiliation{Ohio State University, Columbus, Ohio 43210, USA}
\affiliation{Old Dominion University, Norfolk, VA, 23529, USA}
\affiliation{Panjab University, Chandigarh 160014, India}
\affiliation{Pennsylvania State University, University Park, Pennsylvania 16802, USA}
\affiliation{Institute of High Energy Physics, Protvino, Russia}
\affiliation{Purdue University, West Lafayette, Indiana 47907, USA}
\affiliation{Pusan National University, Pusan, Republic of Korea}
\affiliation{University of Rajasthan, Jaipur 302004, India}
\affiliation{Rice University, Houston, Texas 77251, USA}
\affiliation{Universidade de Sao Paulo, Sao Paulo, Brazil}
\affiliation{University of Science \& Technology of China, Hefei 230026, China}
\affiliation{Shandong University, Jinan, Shandong 250100, China}
\affiliation{Shanghai Institute of Applied Physics, Shanghai 201800, China}
\affiliation{SUBATECH, Nantes, France}
\affiliation{Texas A\&M University, College Station, Texas 77843, USA}
\affiliation{University of Texas, Austin, Texas 78712, USA}
\affiliation{Tsinghua University, Beijing 100084, China}
\affiliation{United States Naval Academy, Annapolis, MD 21402, USA}
\affiliation{Valparaiso University, Valparaiso, Indiana 46383, USA}
\affiliation{Variable Energy Cyclotron Centre, Kolkata 700064, India}
\affiliation{Warsaw University of Technology, Warsaw, Poland}
\affiliation{University of Washington, Seattle, Washington 98195, USA}
\affiliation{Wayne State University, Detroit, Michigan 48201, USA}
\affiliation{Institute of Particle Physics, CCNU (HZNU), Wuhan 430079, China}
\affiliation{Yale University, New Haven, Connecticut 06520, USA}
\affiliation{University of Zagreb, Zagreb, HR-10002, Croatia}

\author{M.~M.~Aggarwal}\affiliation{Panjab University, Chandigarh 160014, India}
\author{Z.~Ahammed}\affiliation{Lawrence Berkeley National Laboratory, Berkeley, California 94720, USA}
\author{A.~V.~Alakhverdyants}\affiliation{Joint Institute for Nuclear Research, Dubna, 141 980, Russia}
\author{I.~Alekseev~~}\affiliation{Alikhanov Institute for Theoretical and Experimental Physics, Moscow, Russia}
\author{J.~Alford}\affiliation{Kent State University, Kent, Ohio 44242, USA}
\author{B.~D.~Anderson}\affiliation{Kent State University, Kent, Ohio 44242, USA}
\author{C.~D.~Anson}\affiliation{Ohio State University, Columbus, Ohio 43210, USA}
\author{D.~Arkhipkin}\affiliation{Brookhaven National Laboratory, Upton, New York 11973, USA}
\author{G.~S.~Averichev}\affiliation{Joint Institute for Nuclear Research, Dubna, 141 980, Russia}
\author{J.~Balewski}\affiliation{Massachusetts Institute of Technology, Cambridge, MA 02139-4307, USA}
% ...

\collaboration{STAR Collaboration}\noaffiliation

\begin{abstract}
Identified charged pion, kaon, and proton spectra are used to
explore the system size dependence of bulk freeze-out properties
in Cu+Cu collisions at \snn=200 and 62.4~GeV.  The data are
studied with hydrodynamically-motivated Blast-wave and statistical
model frameworks in order to characterize the freeze-out properties
of the system.  The dependence of freeze-out parameters on beam
energy and collision centrality 
is discussed. Using the existing results from Au+Au and \pp~collisions,
the dependence of freeze-out parameters on the system size is also
explored.
This multi-dimensional systematic study furthers our understanding
of the QCD phase diagram revealing the importance of the initial
geometrical overlap of the colliding ions. 
The analysis of Cu+Cu collisions, which expands the system size
dependence studies from Au+Au data with detailed measurements in the
smaller system, shows that the bulk freeze-out properties of charged
particles studied here scale with the total charged particle
multiplicity at mid-rapidity, suggesting the relevance of initial
state effects.

\end{abstract}


\maketitle

\section{INTRODUCTION}

An experimental study of relativistic heavy ion collisions augments
our understanding of the QCD phase diagram~\cite{cite:QCD_Diagram}.
The high energy density reached in such collisions at the Relativistic
Heavy Ion Collider (RHIC) is believed to result in a novel state of
hot and dense matter with properties strikingly different from that
of a hadron gas or ordinary nuclear matter~\cite{cite:STARWitePaper}.  

The bulk properties of particle production are studied using
identified particle spectra at low momentum. Model-dependent
interpretations of the measured data provide insight into the
complex dynamics of the collision and further explore the QCD
phases through which the collision evolves.  The dense system
formed in the early stages of the collision continuously
expands and cools, until kinetic freeze-out, beyond which the
particles stream freely into the detector.  Through
measurements of species abundances and transverse momentum
distributions, information about the final stages of the
collision evolution at  chemical and kinetic freeze-out can
be inferred.

The relative particle abundances and spectral shapes discussed
here were tested within the frameworks of statistical (chemical
equilibrium)~\cite{cite:StatModel} and
Blast-wave~\cite{cite:BlastWaveModel} models.  In the chemical
equilibrium model, particle abundances relative to the total system
volume (assumed to be the same for all particle species) are
described by system temperature at freeze-out,  the baryon and
strangeness chemical potentials, and the strangeness
suppression factor. The Blast-wave model describes spectral shapes
assuming a locally thermalized source with a common  transverse
flow velocity field.  The results from Au+Au collisions at
\snn=200 and at
62.4~GeV~\cite{cite:200spectra,cite:62spectra,cite:LongPRC} have
shown that the chemical freeze-out temperature, $T_{\rm ch}$, has
little dependence on centrality whereas the kinetic freeze-out
temperature, $T_{\rm kin}$, decreases with increasing centrality of the
collision. Further, the radial flow velocity, $\beta$, increases
with increasing centrality.  The observed changes in $T_{\rm kin}$ and
$\beta$ with  centrality are consistent with higher energy and
pressure in the initial state for more central events.  On the
other hand, the centrality independence of the extracted chemical
freeze-out temperature indicates that even for different initial
conditions, collisions always evolve to the same chemical
freeze-out. Moreover, the value for the chemical freeze-out
temperature in Au+Au is close to the critical temperature,
predicted by some lattice QCD calculations~\cite{Tcrit}.  This
suggests that chemical freeze-out  coincides with hadronization
and therefore  $T_{\rm ch}$  provides a lower limit estimate
for the temperature of the prehadronic state~\cite{cite:Olgaposter}.
The systematic behavior of the kinetic freeze-out properties with
charged hadron multiplicity appears to follow the same trend for
all energies and systems at RHIC~\cite{cite:200spectra,cite:62spectra}.
In this paper, the systematic studies of the QCD phase diagram
from heavy ion collisions are enriched by the addition of new
RHIC data from Cu+Cu collisions at \snn=200 and 62.4~GeV.  

\section{THE STAR EXPERIMENT}
The Cu+Cu data presented here were collected by the STAR experiment
during the RHIC 2005 run.  Copper nuclei ($^{63}$Cu) were collided
at \snn=200 and 62.4~GeV.  Data were recorded with a minimum bias
trigger obtained from  the Beam-Beam Cherenkov counters~\cite{cite:BBCs}
coupled with information from the Zero-Degree Calorimeters~\cite{cite:ZDCs}.
This trigger is found to be sensitive to the top $\sim$85\% of
the inelastic cross-section.  The data studied here correspond to
the top 60\% of the inelastic cross-section (minimum bias) where
little or no inefficiency of the triggering or vertex reconstruction
is found.  These 0-60\% minimum bias events, with 24~M and 10~M
events recorded at 200 and 62.4~GeV, respectively,  were divided
into six centrality bins each corresponding to a 10\% interval of
the geometric cross section.  Transverse mass distributions for
charged pions, charged kaons, protons and antiprotons, previously
reported by STAR for \snn=200 GeV $pp$ collisions and Au+Au
collisions  at 62.4 and
200~GeV~\cite{cite:200spectra,cite:62spectra,cite:LongPRC}, are used
for comparison.

\begin{figure*}[!ht]
\includegraphics[width=0.475\textwidth]{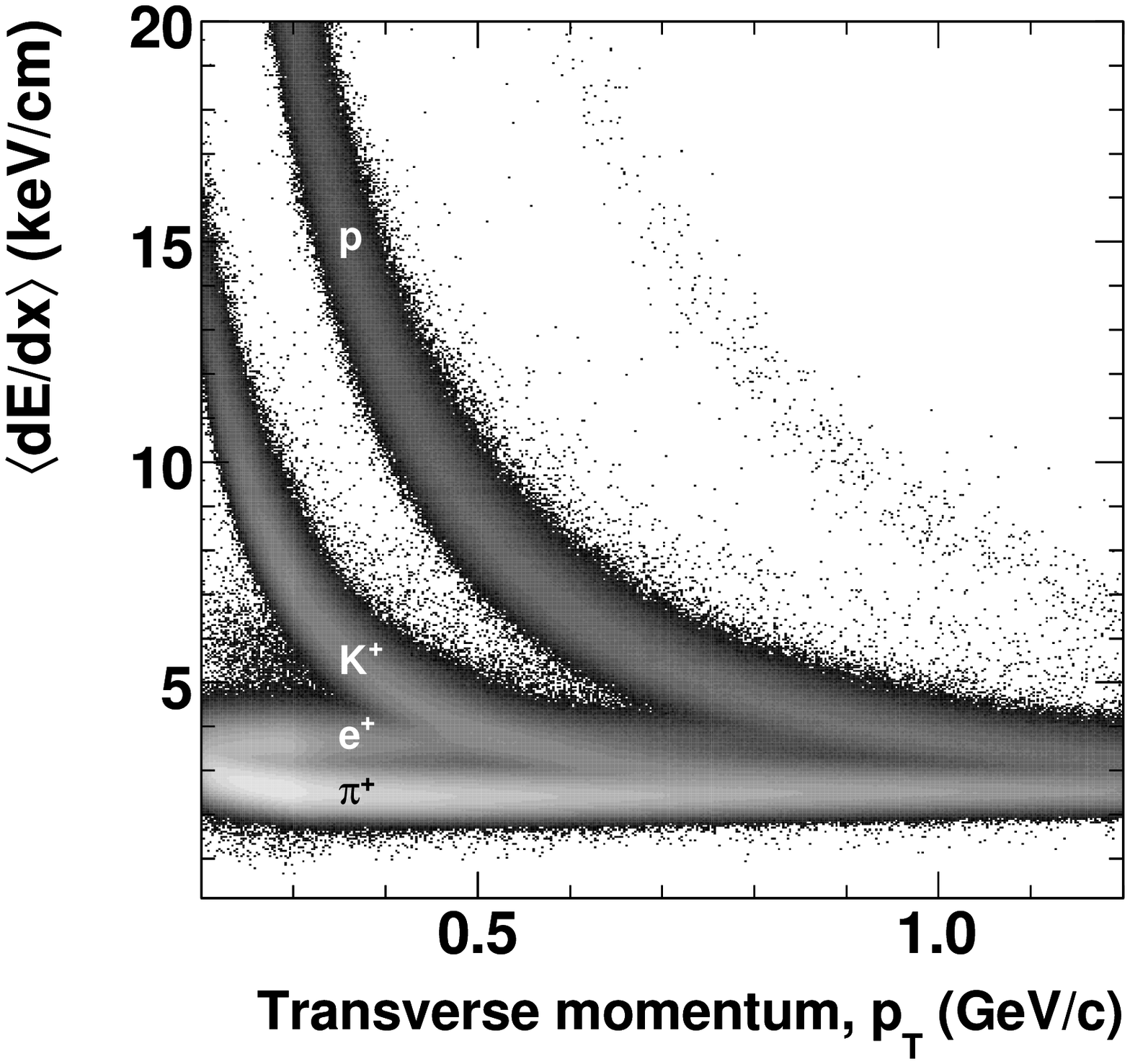}
\includegraphics[width=0.475\textwidth]{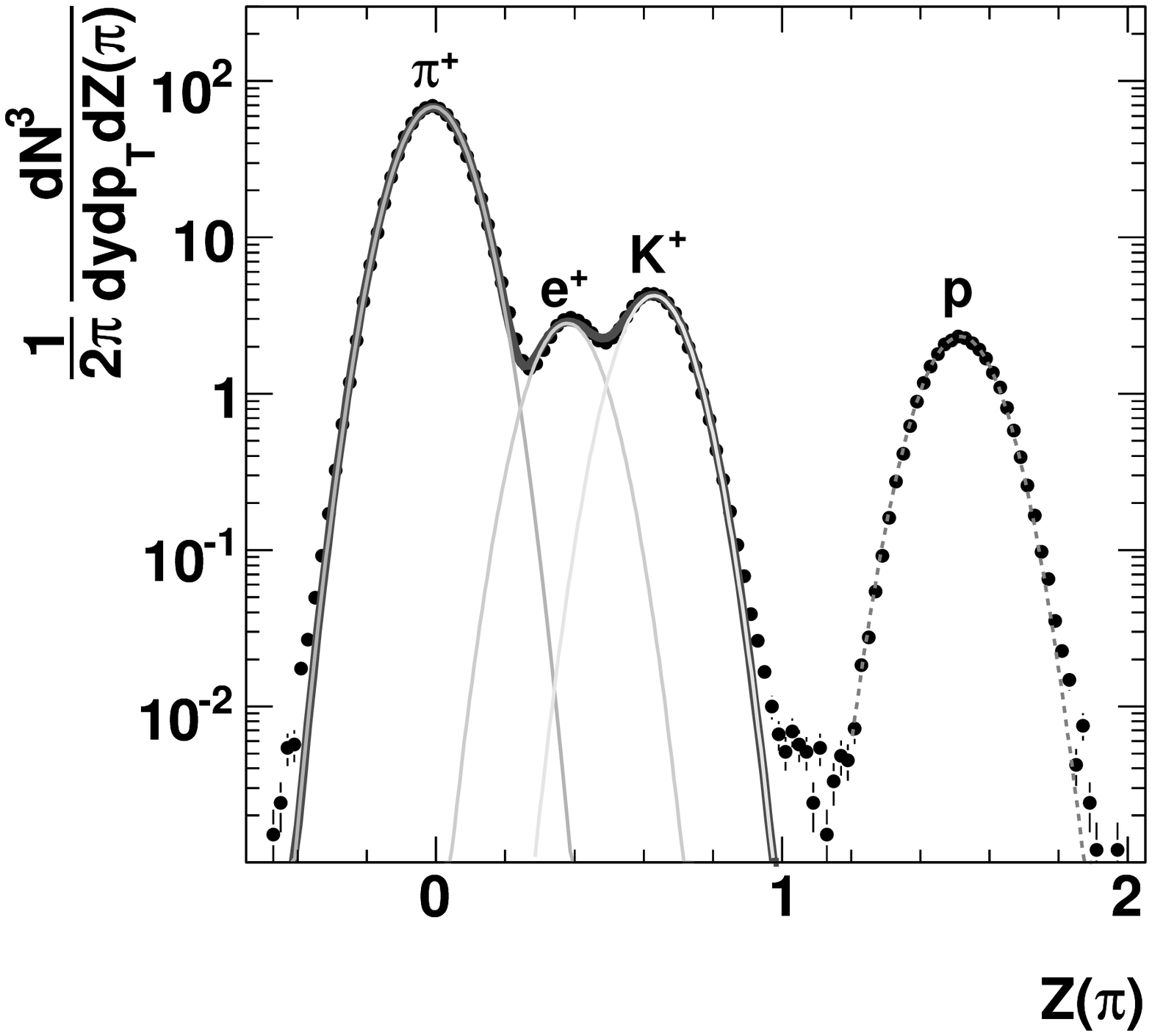}
\caption{\label{fig:dedx}  The left panel shows the truncated mean
ionization energy loss ($\langle dE/dx \rangle$) in the TPC as a 
function of transverse momentum for positively charged tracks from
200~GeV Cu+Cu collisions.  The right panel shows $Z(\pi)$, the
logarithm of the measured $\langle dE/dx \rangle$ divided by the
theoretical expectation for energy loss of charged pions, for
$0.40 < \pT < 0.55$~GeV/$c$.  Also shown is an example four-Gaussian
fit that is used to extract the raw yields for different species.}
\end{figure*}

\begin{figure}[!ht]
\includegraphics[width=0.475\textwidth]{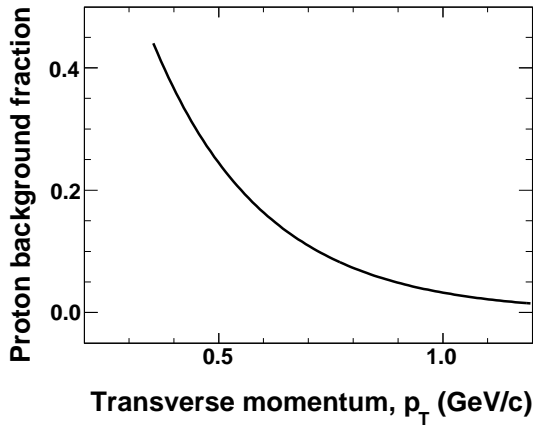}
\vspace*{-0.3cm}
\caption{\label{fig:pBack} Estimated fraction of background protons in
the raw proton sample as function of transverse momentum, for the most
central \snn=200~GeV Cu+Cu collisions.  No strong centrality or energy
dependence for this correction was observed for all Cu+Cu data available.}
\end{figure}

\begin{figure*}[!ht]
\includegraphics[width=0.95\textwidth]{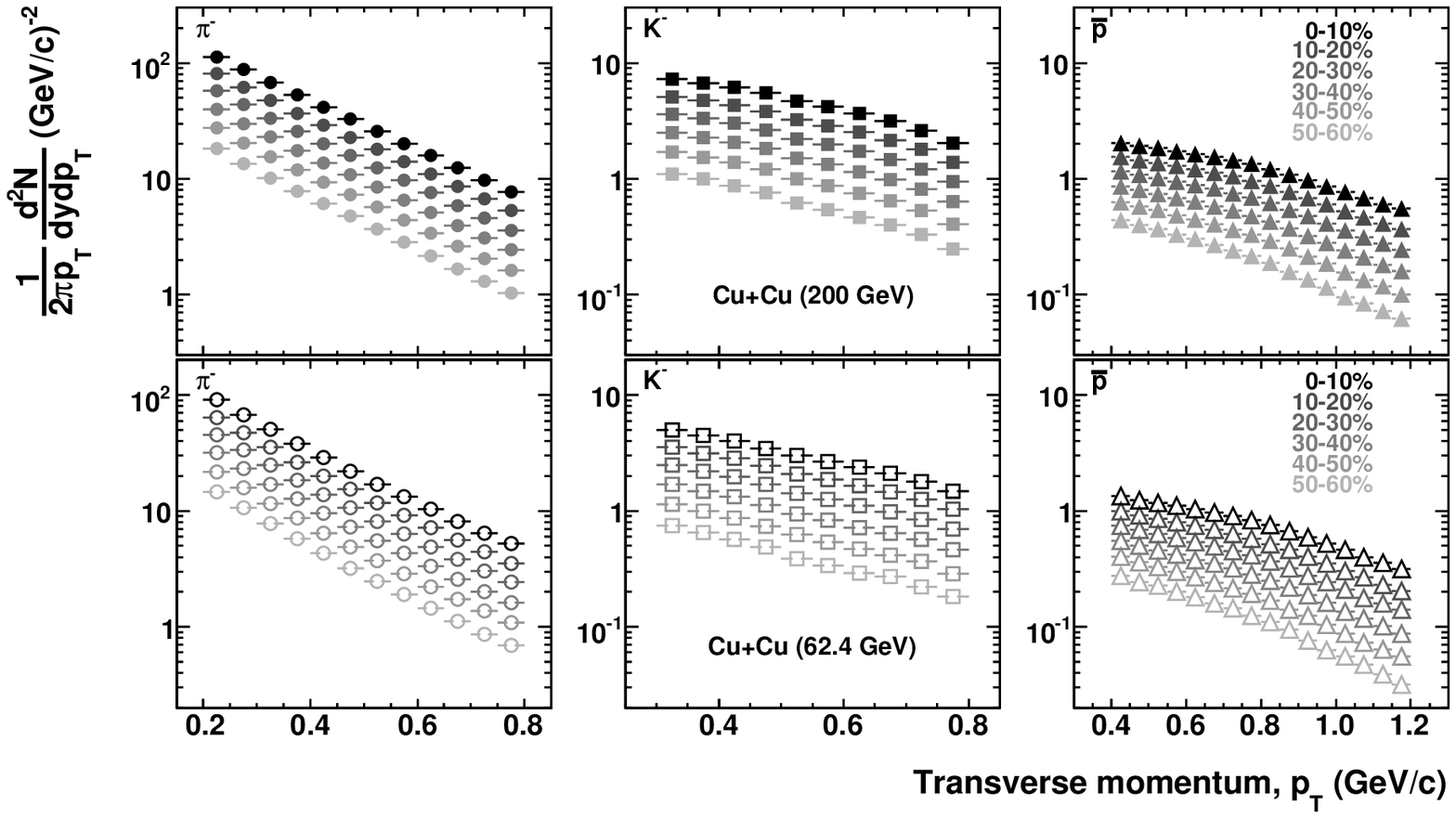}
\vspace{-0.5cm}
\caption{\label{fig:SpectraMinus}The top row shows negatively charged pion
(leftmost column), kaon (center) and anti-proton (right) spectra from Cu+Cu
collisions at \snn=200~GeV.  Six centrality classes are shown as dark
(central 0-10\%) to light (50-60\%) shades.  
Spectra for 62.4~GeV Cu+Cu are shown on the bottom row.  Statistical and
systematic errors (which do not exceed 7\%) are smaller than the symbol size.}
\end{figure*}

\begin{figure*}[!ht]
\includegraphics[width=0.95\textwidth]{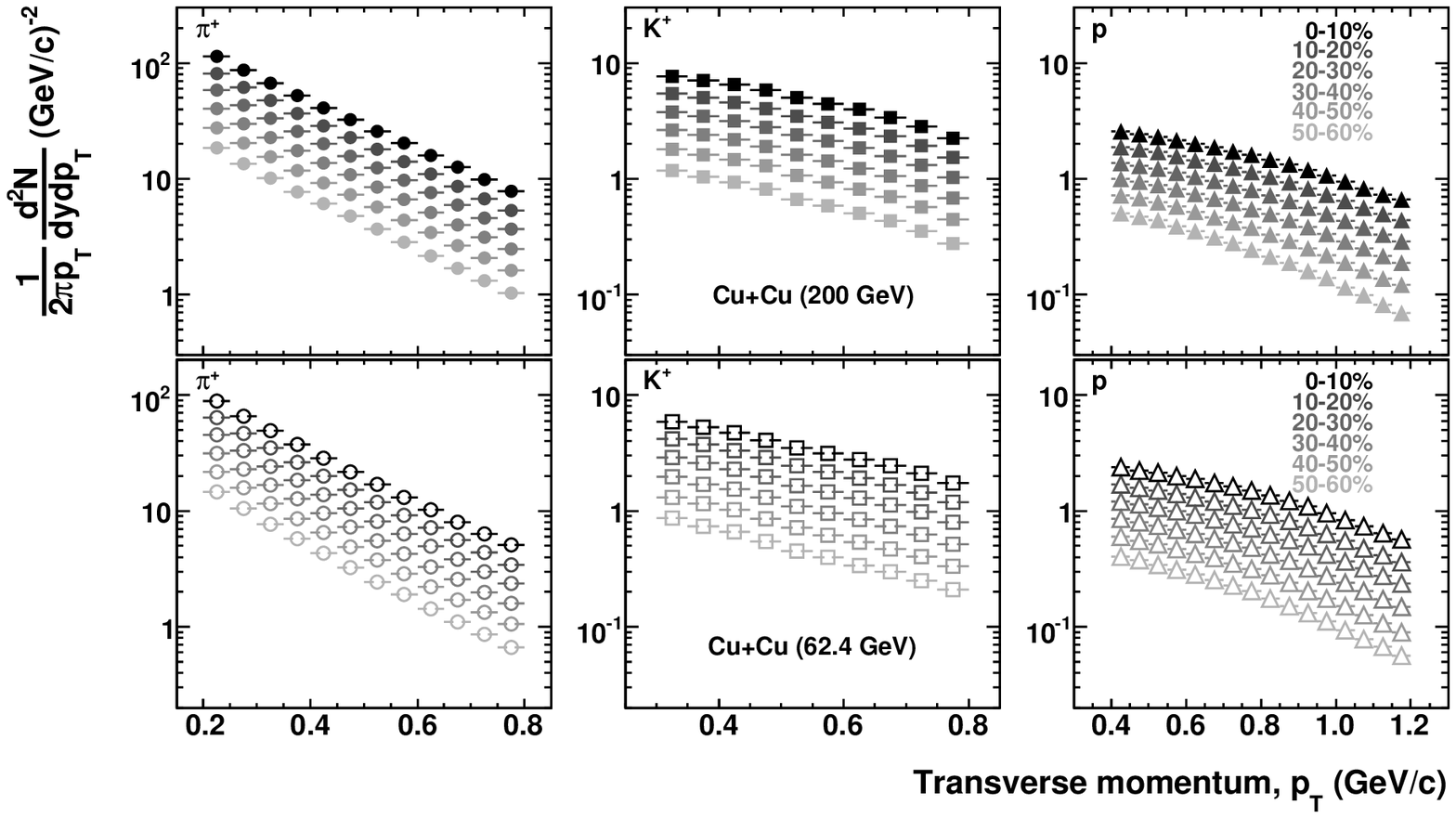}
\vspace{-0.8cm}
\caption{\label{fig:SpectraPlus}The top row shows positively charged pion
(leftmost column), kaon (center) and proton (right) spectra from Cu+Cu
collisions at \snn=200~GeV.  Six centrality classes are shown as dark
(central 0-10\%) to light (50-60\%) shades.  
Spectra for 62.4~GeV Cu+Cu are shown on the bottom row.  Statistical and
systematic errors (which do not exceed 7\%) are smaller than the symbol size.}
\end{figure*}
 
\begin{figure*}[!ht]
\includegraphics[width=0.98\textwidth]{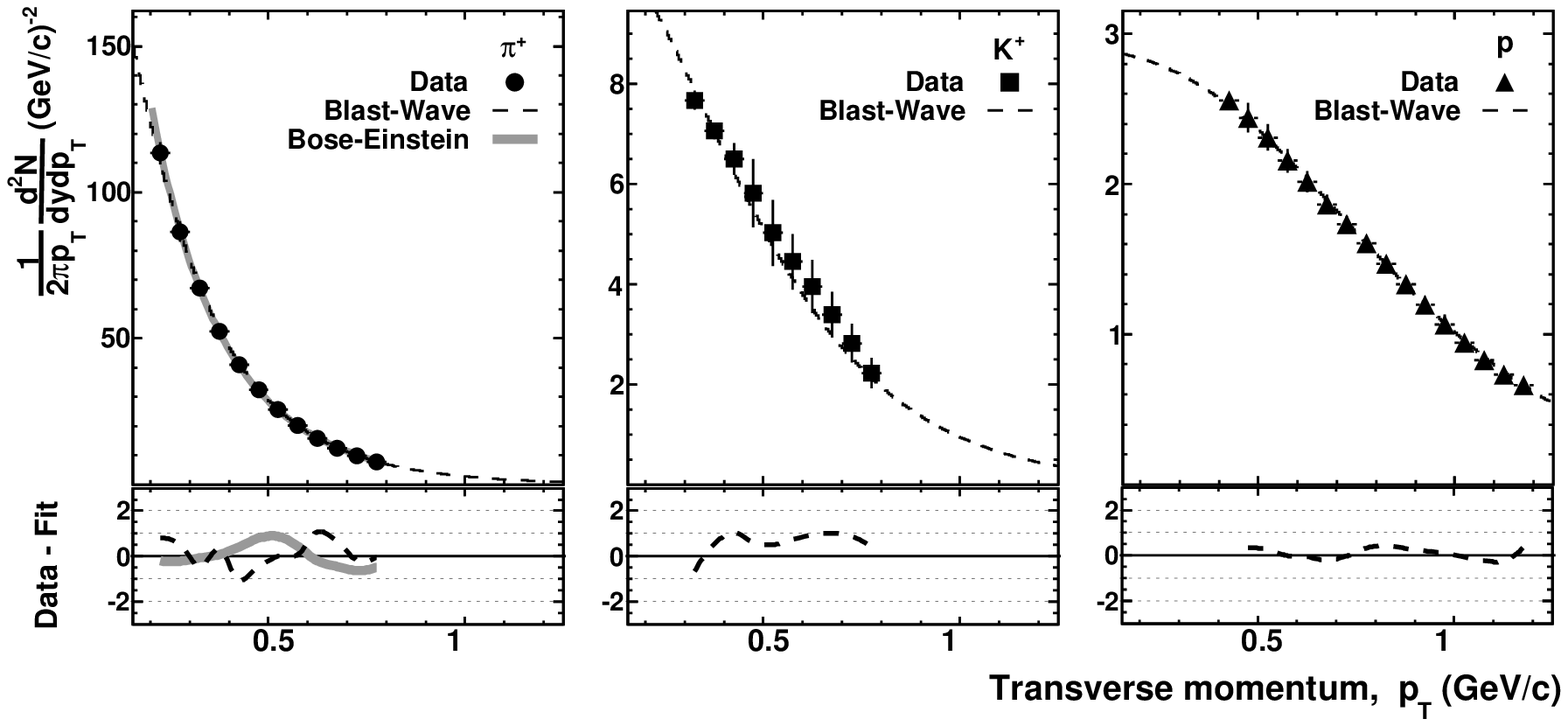}
\caption{\label{fig:SpecFits}
Comparison between the 10\% most central Cu+Cu collision data (symbols)
at \snn=200~GeV  and the corresponding Blast-wave model fit (dashed
line) to $\pi^{+}$ (left), K$^{+}$ (center) and proton (right) spectra
-- note fit is performed simultaneously across species.  The pion data
points below \pT~=~0.5~GeV/$c$ were not included in the Blast-wave fits
to reduce the effect of resonance decays. 
A Bose-Einstein fit  to the pion spectra over the entire
fiducial range is also shown.  The lower panels illustrate the quality
of the fits by showing the difference between the measured points and
the fit expressed as the number of standard deviations.}
\end{figure*} 

The STAR Time Projection Chamber (TPC)~\cite{cite:STAR_TPC} tracks
particle trajectories over a wide range of momentum at mid-rapidity
($|\eta| < 1.8$).  The particle identification at low-\pT~uses
measurements of truncated mean ionization energy loss,
$\langle dE/dx \rangle$, of the charged particles traversing the
TPC.  Particles of different mass show distinct patterns in the
$\langle dE/dx \rangle$ dependence, as shown in Fig.~\ref{fig:dedx},
left panel.  This allows statistical separation of pions and kaons
in the momentum range $0.25 < \pT < 0.80$~GeV/{\em c} at mid-rapidity
($|y| < 0.1$), and of protons and anti-protons from other species in
the range $0.40 < \pT < 1.20$~GeV/{\em c}.

The momentum measurement is given by the curvature of the particle
trajectories as they pass through the 0.5~T magnetic field of the
STAR detector.  
To ensure optimal $dE/dx$ resolution, only primary tracks, with {\em dca}
(distance of closest approach between the particle trajectory and the
event vertex) less than 3~cm, and at least 25 out of 45 possible fit
points are used in this analysis.
Particle identification  at mid-rapidity ($|y| < 0.1$)
is achieved by fits to the $Z$-variable, defined as a logarithm of
$\langle dE/dx \rangle$ divided by the theoretically expected value
for each particle type, given by Bethe-Bloch formula~\cite{cite:BB}.
This new variable is introduced to remove the
strong \pT~dependence at low momenta.  Such a normalized distribution
is created for a given particle and centrality and is divided into
narrow transverse momentum slices (width $\Delta$\pT~=~50~MeV/{\em c}).
These momentum projections are fit with a combined four-Gaussian
function, one for each of the particle species of a given charge:
$\pi$, K, $p$ and $e$.  
The integral of each Gaussian provides the raw yield at each momentum.
This procedure is repeated for each particle species in order to assign
the correct rapidity for each track, using the mass of the particle.
Thus, fits to the auxiliary particles in each distribution (for example
K, $p$ and $e$ for $\pi$ analysis) are used only to estimate the
contamination when bands  overlap. The right panel of Fig.~\ref{fig:dedx}
shows an example for pion yield extraction for one momentum slice. For
more details see Ref.~\cite{cite:LongPRC}.

The raw yields extracted from each of the four-Gaussian fits are then
corrected for detector acceptance, single-track reconstruction
efficiencies, and other effects as discussed below. To determine the
correction factors, simulated tracks were embedded into real data on
the raw signal level and run through the standard reconstruction chain.
The estimated single-track reconstruction efficiency is about $80\%$
for $\pi^{\pm}$ in Cu+Cu collisions and exhibits a small centrality
and $p_T$ dependence.  The \pT-spectrum has also been corrected for the
energy loss
by multiple scattering beyond that for pions which is calculated
during reconstruction.  This affects the reconstructed momentum at
low values. The maximum value of this  mass-dependent correction to
the measured \pT~value for ${\rm K}^{\pm}$ and $p$($\overline{p}$) was found
to be $2\%$ and $3\%$, respectively, for the lowest measured \pT~bin.  
An additional correction for the background contamination in the proton
sample is made.  The background protons arise predominantly from
secondary interactions in the beam pipe and detector material
(knock-out protons). It is estimated from data to be about 40\% at
\pT~=~400~MeV/{\em c}, diminishing to near zero at \pT~=~1~GeV/{\em c},
as shown in Fig.~\ref{fig:pBack}.
To estimate this correction factor we compare the distribution of proton
{\em dca} 
to that of the anti-protons (see
Ref.~\cite{cite:LongPRC} for more details).  In the measured {\em dca}
region, the integral difference between protons and anti-protons (after
normalization by the anti-proton to proton ratio) is considered to be
the background contribution to the proton yield.
Pion yields are additionally corrected for feed-down contributions from
weakly decaying particles, muon contamination, and background pions
from detector material. This correction is found to decrease from about
$15\%$ at 0.3~GeV/$c$ to about $5\%$ at 1~GeV/{\em c}.
The (anti)protons presented in this paper are inclusive measurements
(not corrected for weak decays). It has been found in previous studies
that the analysis cuts used for the low-\pT~identified proton studies
($dca < 3$~cm) reject only a negligible fraction of daughter protons from
the hyperon decays~\cite{cite:200spectra}. Therefore, our  sample 
reflects the total baryon production in the collision. Earlier Au+Au
studies~\cite{cite:200spectra} and preliminary Lambda-hyperon spectra
from Cu+Cu collisions~\cite{cite:AntStrange} indicate that the
freeze-out spectral shapes are similar for $\Lambda$s and protons,
resulting in similar spectra shapes for primary and feed-down protons.
The fraction of the weak-decay feed-down protons is estimated to be
about $30\%$~\cite{cite:JuneWWND08}.

This analysis technique is used to obtain the low-\pT~particle spectra
for all centrality bins at both 200 and 62.4~GeV center-of-mass
energies and for the Cu+Cu and Au+Au colliding systems.  Additional
technical details on  the analysis  and applied corrections can be
found in Refs.~\cite{cite:200spectra,cite:62spectra} with a thorough
overview in Ref.~\cite{cite:LongPRC}.

\section{RESULTS}

The transverse momentum spectra are shown in
Figs.~\ref{fig:SpectraMinus}~and~\ref{fig:SpectraPlus} for $\pi^{\pm}$
(leftmost column), K$^{\pm}$ (center) and (anti)protons (right) in
Cu+Cu collisions. The top row presents the data for \snn=200~GeV, whilst 
data for 62.4~GeV are shown in the bottom row.  The symbol shades represent
different centrality bins.  The particle and anti-particle spectral
shapes are similar for all species in each centrality bin. At both
collision energies a mass dependence is observed in the slope of the
particle spectra.  Due to the large number of events recorded and good
tracking efficiency, the statistical errors are less than 1\%.  The
systematic uncertainties are similar to those determined in prior
analyses of low-\pT~spectra in Au+Au collisions~\cite{cite:LongPRC}.
Systematic errors are divided into two classes: point-to-point and
scale uncertainties.  The overall scale uncertainty, mostly due to the
embedding procedure for the single-track reconstruction efficiency, is
estimated to be 5\%  for all particle species.  Point-to-point
uncertainties are determined for each \pT~bin and particle species.
For pions and kaons, this error is evaluated to be less than 7\% and
13\%, respectively.  These maximal errors represent \pT~bins where a
significant $\langle dE/dx \rangle$ overlap occurs between $\pi^{\pm}$,
K$^{\pm}$ or $e^{\pm}$.  For protons and anti-protons, the maximum
error is 5\%.  At low-\pT, the proton uncertainty is greater than that
for anti-protons (4.0\% versus 1.3\%, respectively, at 
\pT~=~400-450~MeV/$c$) owing to the additional uncertainty from the
proton's background.  The uncertainty due to the background decreases
rapidly from 3.7\% at \pT~=~400-450~MeV/$c$ to 1.5\% for $\pT > 1$~GeV/$c$.

For the anti-particle to particle yield ratios, systematic errors are
much reduced due to a cancellation of the efficiency uncertainties
and a partial cancellation of extrapolation uncertainties, as described
above.  A  systematic uncertainty of 2\%, 3\%, and 5\% is assigned to
$\pi^{-}/\pi^{+}$, K$^{-}$/K$^{+}$, $\overline{p}/p$, respectively.

We further fit the obtained \pT~distributions  to extract system
properties at different stages of the collision evolution.  The first
fit to the data probes collision properties at kinetic freeze-out.
Here, a Blast-wave model~\cite{cite:BlastWaveModel} is used to
simultaneously fit the $\pi^{\pm}$, K$^{\pm}$ and (anti)proton spectra
at a given centrality.  This fit provides a good description of the
spectra shapes, as illustrated in Fig.~\ref{fig:SpecFits} with results
from most central 200~GeV Cu+Cu data.  The $\pi^{\pm}$ data points for
$\pT < 0.5$~GeV/$c$ are excluded from the Blast-wave fits to reduce the
effects of resonance decay contributions as done in previous
works~\cite{cite:200spectra,cite:62spectra,cite:LongPRC}.  Including
this  low-\pT~region in the fit leads to a poorer description of proton
and kaon shapes, however the resultant modification of the extracted
parameters remains well within  their systematic uncertainty.  The
freeze-out parameters obtained from this  model are discussed later.
Also shown in Fig.~\ref{fig:SpecFits} are Bose-Einstein
($\propto 1/(\exp{\frac{\mT}{T}}-1)$) fits to the $\pi^{\pm}$, which
provide a slightly better description of these data. For evaluation
of the systematic uncertainties from extrapolation, $m_{T}$-exponential
($\propto 1/\exp{\frac{\mT}{T}}$) and Boltzmann
($\propto \mT/\exp{\frac{\mT}{T}}$) fits are also used in the
analysis (for more details see Ref.~\cite{cite:LongPRC}).

\begin{figure}[!t]
\includegraphics[width=0.475\textwidth]{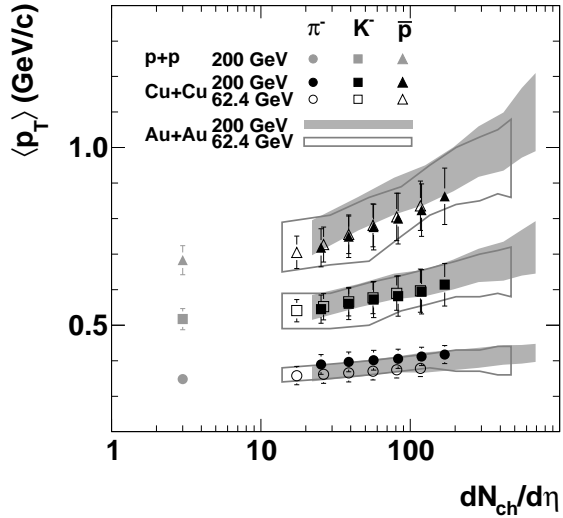}
\caption{\label{fig:MeanPt} Mean transverse momentum as a function of
charged hadron multiplicity at mid-rapidity for pions, kaons and
anti-protons.  For comparison, the mean-\pT~values for Au+Au data are
shown as bands.  Open (closed) symbols/bands depict data at
\snn=200~GeV (62.4~GeV).  Error-bars represent statistical and systematic
uncertainties added in quadrature. }
\end{figure}

\begin{figure}[!t]
\includegraphics[width=0.475\textwidth]{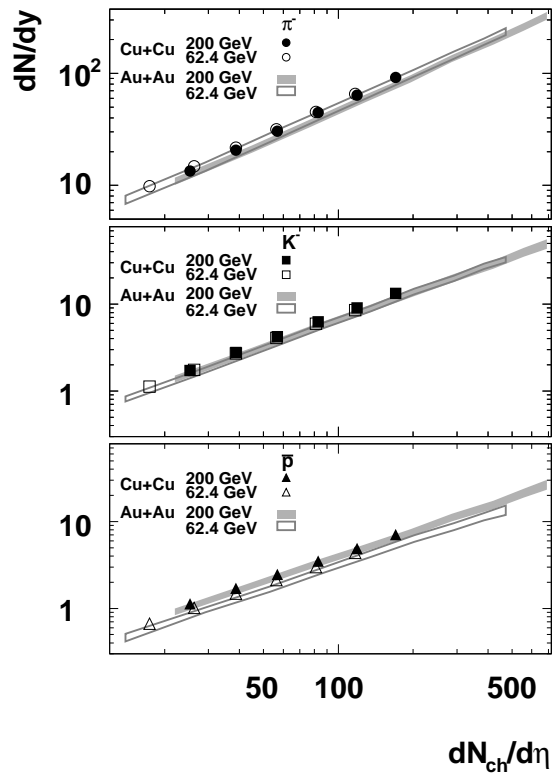}
\caption{\label{fig:dN/dy} Integrated yields at mid-rapidity for pions, kaons and
anti-protons  as a function of the charged particle density (\nch),
which is used as a measure of the centrality.  For comparison Au+Au
data are shown as bands.  Filled (open) points/bands depict data at
\snn=200~GeV (62.4~GeV).  Error-bars represent statistical and systematic
uncertainties added in quadrature.}  
\end{figure}

The particle mean-\pT~and total particle yields at mid-rapidity
($|y| < 0.1$) are shown in Fig.~\ref{fig:MeanPt} and Fig.~\ref{fig:dN/dy},
respectively. The values presented for kaons and (anti)protons are
determined from the measured spectra points, extrapolated outside the
fiducial range using Blast-wave fits discussed above. Similarly, a
combination of the measured data-points and extrapolation from
Bose-Einstein fits is used for the pions.  The measured fraction of
the total yield is found to be 62\% for $\pi^{\pm}$, 58\% for K$^{\pm}$
and 65\% for (anti)protons  for the most central 200 GeV data; these
fractions are slightly higher in other centrality bins and at lower
energy~\cite{cite:STAR9GeV}.  The systematic uncertainty on $dN/dy$ and
mean-\pT, shown in
the figures, includes the extrapolation uncertainty evaluated by means
of various model fits  mentioned earlier. Overall, these are estimated
to be near 15\% of the yields outside the fiducial range for pions and
kaons and 15\%-25\% of the extrapolated yields for protons and
anti-protons, depending on centrality.

We also determine the total charged hadron production per unit of
pseudo-rapidity, \nch, at mid-rapidity.  The total particle yield at
mid-rapidity for each species, obtained by extrapolating the fits to
the measured spectrum in the momentum range outside our fiducial
coverage,  was corrected for the Jacobian transformation
\nchy$\rightarrow$\nch. The sum of the total charged pion, kaon and
(anti)proton yields was then corrected for  the feed-down of weakly
decaying neutral strange particles, providing the estimate of
primordial charged hadron yield at mid-rapidity.  A complementary
method was also used,  integrating over the charged hadron spectra
corrected for efficiency, feed-down and the Jacobian transformation,
and yielded consistent results.

The mean-\pT~of each particle species ($\pi$, K, $p$) increases with
the number of charged hadrons at mid-rapidity \nch,  as shown in
Fig.~\ref{fig:MeanPt}.  Moreover, the mean-\pT~for each particle
species appears to scale with \nch~at mid-rapidity, and to be
independent of the colliding system and the center-of-mass energy.  
The particle yields show the same systematic scaling features with
\nch~as mean-\pT~across system and collision energy, see
Fig.~\ref{fig:dN/dy}.  In this logarithmic representation the particle
yields for each species appear to increase linearly with multiplicity,
with Cu+Cu matching the Au+Au data at similar values of \nch. When
shown on a linear scale, the integrated yields exhibit a near-linear
dependence with \nch.  The logarithmic scale for both axes used here
preserves the apparent linear dependencies whilst better illustrating
the lower multiplicity Cu+Cu data.  For the detailed features we
investigate the relative particle production  in the following.

\begin{figure}[!t]
\includegraphics[width=0.475\textwidth]{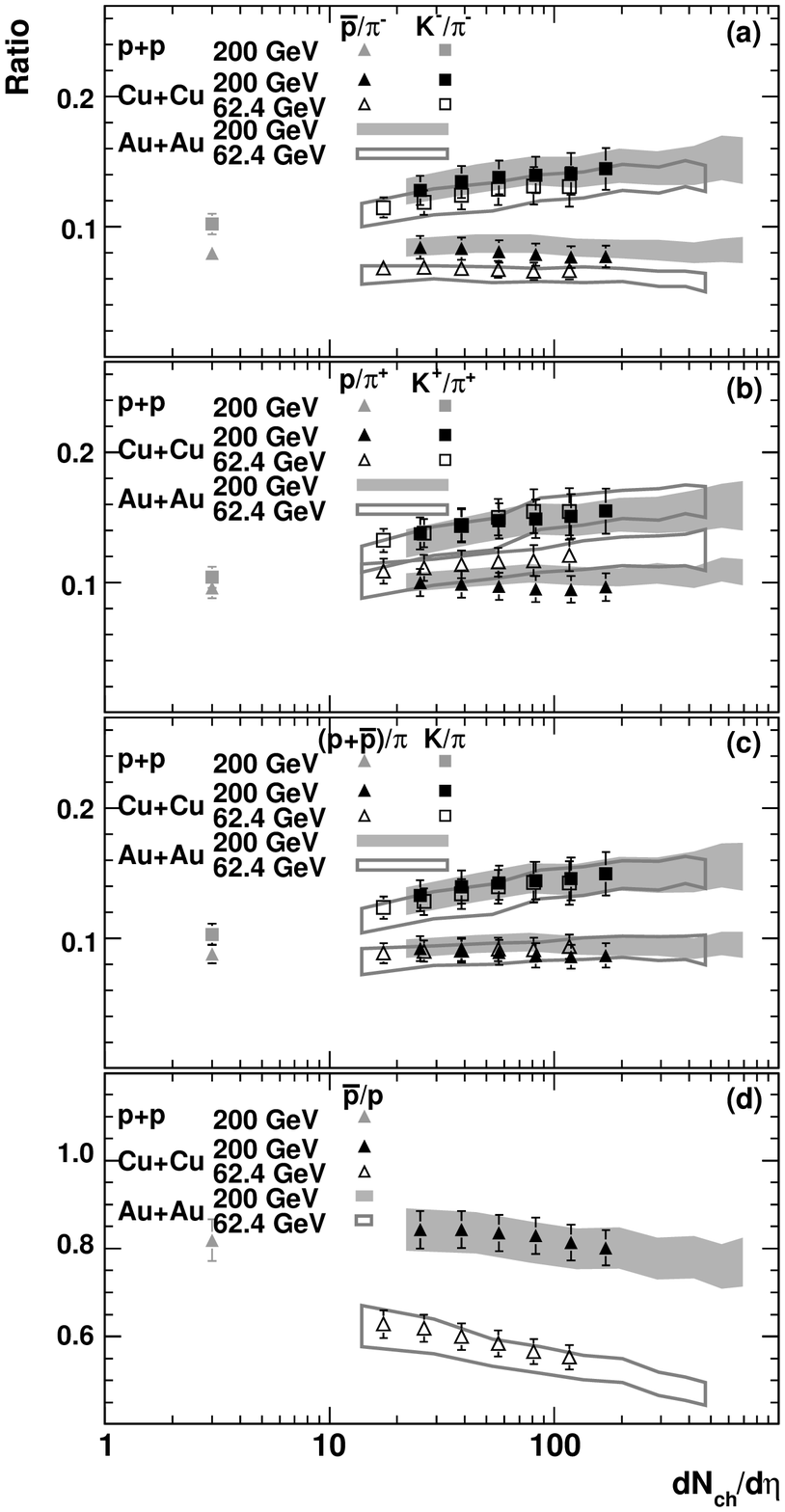}
\caption{\label{fig:Ratios} Integrated particle yield ratios at
\snn=200~GeV (closed symbols) and 62.4~GeV (open) for Cu+Cu (black)
and Au+Au collisions (grey bands) versus \nch~at mid-rapidity. Error-bars
represent statistical and systematic uncertainties added in quadrature.}
\end{figure}

The relative abundances of particles provide an important insight
into the chemical properties of the system.  The relative kaon yield
reflects the strangeness production  in the collision, whereas proton
with respect to pion production is dependent on the baryon production
and transport.  Figure~\ref{fig:Ratios}($a$) shows the ratios for the
negatively charged particles, $\overline{p}/\pi^{-}$ and ${\rm K}^{-}/\pi^{-}$,
as a function of \nch, which exhibit similar \nch-scaling behavior at
each collision energy.  
The slight decrease of the values for both ratios seen at the
lower collision energy of
 62.4~GeV is insignificant within
experimental uncertainties.  Figure~\ref{fig:Ratios}($b$) shows the
ratios for positively charged particles, $p/\pi^{+}$ and ${\rm K}^{+}/\pi^{+}$,
which also exhibit a \nch-scaling behavior within the same collision
energy. The beam-energy effect is reversed here as compared to the ratio
of negatively charged particles. Summing over the two charges
(Fig.~\ref{fig:Ratios}($c$)), the corresponding ratios exhibit a common
scaling behavior with \nch, independent of colliding system and collision
energy. The energy dependence of the positive and negative particle
ratios considered separately, points to the effects of baryon transport
to mid-rapidity, which decreases with increasing energy (see also~\cite{cite:STAR9GeV}).

\begin{figure*}[t]
\includegraphics[width=0.85\textwidth]{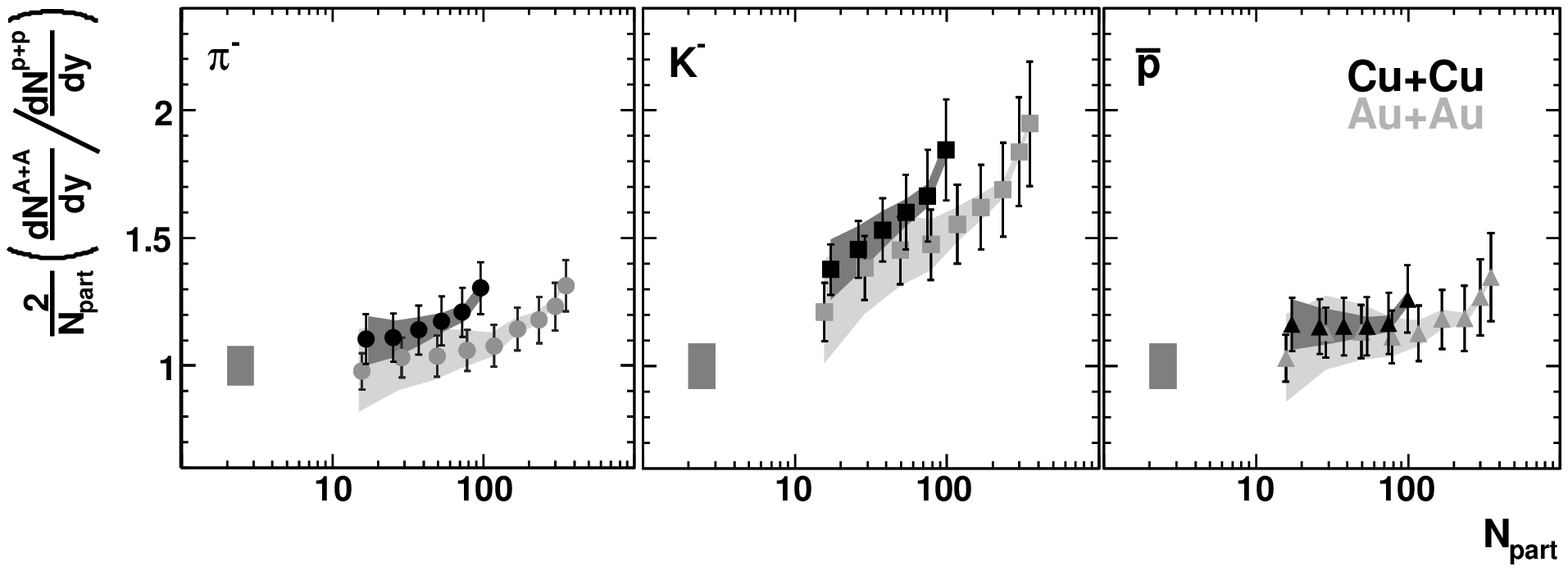}
\caption{\label{fig:Enhancement} Enhancement factors for negatively
charged pions (left), kaons (center), and anti-protons (right) as
function of \npart~in \snn=200  Cu+Cu and Au+Au collisions.
Error-bars represent statistical and systematic uncertainties on the
A+A measurements added in quadrature. The shaded bands depict model
uncertainties on number of participants calculation. The bands on the
left show uncertainties from the $pp$ measurements that are correlated
for all data points. }
\end{figure*}

We further explore the kaon production in Cu+Cu collisions to gain
better insight into production of strange quarks.  The ${\rm K}^{-}/\pi^{-}$
ratio scales with \nch~at both energies and there is no hint of an
additional strangeness enhancement of charged kaons in the smaller
Cu+Cu system compared to the larger Au+Au system. Early works from
SPS energies reported such additional relative strangeness
enhancement in the ${\rm K}/\pi$ ratio for smaller systems, although no
final confirmation of this observation is
available~\cite{cite:SPS_NA49QM02,cite:SPS_NA49PRL}.  The pion and
kaon enhancement factors are compared in Fig.~\ref{fig:Enhancement}.
This factor is defined as the yield per mean number of participating
nucleons (estimated using a Glauber model), \npart,  in heavy-ion
collisions divided by the respective value in \pp~collisions. A
progressive enhancement of kaon production with respect to pions as
a function of collision centrality is evident, as shown earlier by
the K/$\pi$ ratios (Fig.~\ref{fig:Ratios}($a$) and
Fig.~\ref{fig:Ratios}($b$)). A comparison of these enhancement
factors between Cu+Cu and Au+Au data is also shown. The enhancement
factors for kaons do not show universal scaling features with
respect to \npart, and are indeed found to be higher  in Cu+Cu
collisions compared to the Au+Au system. However, these features do
not appear to be unique to kaons. A similar trend is observed in
Fig.~\ref{fig:Enhancement} in the
pion enhancement factors for the two systems. This suggests that the
additional enhancement, seen in the charged kaon  yields, is not
related to strangeness production, but other physics mechanisms, for
example, additional entropy production. It should be noted that while
comparing more spherical central Cu+Cu collisions with semi-peripheral
Au+Au collisions, the initial conditions may not be reflected by
\npart~alone.  

In contrast to pions and kaons, protons show minimal evolution with
centrality and no difference between Cu+Cu and Au+Au systems is
observed. Figure~\ref{fig:Ratios}($d$) illustrates the difference in
anti-proton and proton production across energies.  As observed in
other collision systems, the ratio is found to increase and becomes
closer to unity for higher energy
collisions~\cite{cite:PHOBOS_ppPartRatios}.  The anti-proton to
proton ratio gives information on the amount of baryon transport.
In line with the earlier STAR results, our measurements indicate
that while a finite excess of baryons over anti-baryons is still
present at RHIC energies, $p-\bar{p}$ pair production becomes an
important factor.  Little or no change due to an increase in the
system size (centrality) is apparent in the Cu+Cu data at 200~GeV,
while 62.4~GeV data show a decreasing trend with increasing centrality
for this ratio  for both Cu+Cu and Au+Au data.

\section{FREEZE-OUT PROPERTIES}

The particle yields and their ratios provide further information on the
thermal properties of the system at kinetic and chemical freeze-out.

\subsection{Kinetic Properties}
The completion of all elastic scattering marks the final stage of
collision evolution and could be interpreted as a kinetic freeze-out,
where the particle momentum spectra are fixed.  To quantify this
stage, fits are made simultaneously to the spectra of all particle
species, but independently for each centrality class (see
Fig.~\ref{fig:SpecFits} for example).  The fits used here are based
on the previously discussed Blast-wave model~\cite{cite:BlastWaveModel},
which assumes a radially boosted thermal source.  These
hydrodynamically-motivated fits describe the mass dependence of
particle spectral shapes in terms of the radial flow velocity ($\beta$),
the kinetic freeze-out temperature ($T_{\rm kin}$) and the flow velocity
profile exponent ($n$) at the final freeze-out.  The extracted value for
$n$ is not used to derive any physics interpretation.  The effects from
resonance contributions to the pion spectral shape are reduced by
excluding the  low-\pT~data points (below 0.5~GeV/{\em c}).  To enable a
comparison with earlier results on \pp~and Au+Au
collisions~\cite{cite:200spectra,cite:LongPRC}, the same model and the
same procedures for the fits are adopted, thereby avoiding any possible
systematic bias.

\begin{figure*}[t]
\includegraphics[width=0.95\textwidth]{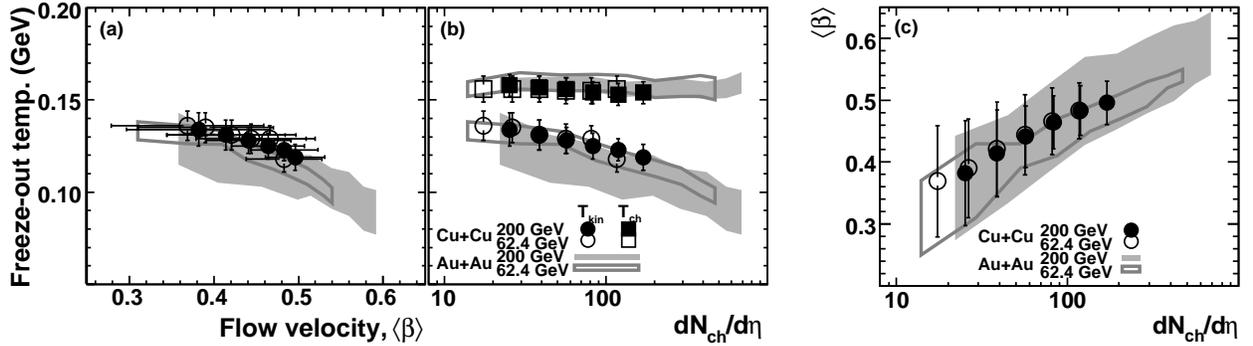}
\caption{\label{fig:KineticFO} Comparison of kinetic freeze-out
properties obtained from fits to Cu+Cu (symbols) and Au+Au (bands)
collision data at \snn=200 (closed symbols/bands) and 62.4~GeV (open).
The kinetic freeze-out temperature, $T_{\rm kin}$, is shown versus flow
velocity, $\beta$, and multiplicity in panels (a) and (b) respectively
(more central collisions are to the right side of each plot).
Panel (b) also shows the multiplicity dependence of the chemical
freeze-out temperature, $T_{\rm ch}$, (square symbols).  Panel (c) shows
the multiplicity dependence of the average radial flow velocity.}
\end{figure*}

\begin{figure*}[t]
\includegraphics[width=0.95\textwidth]{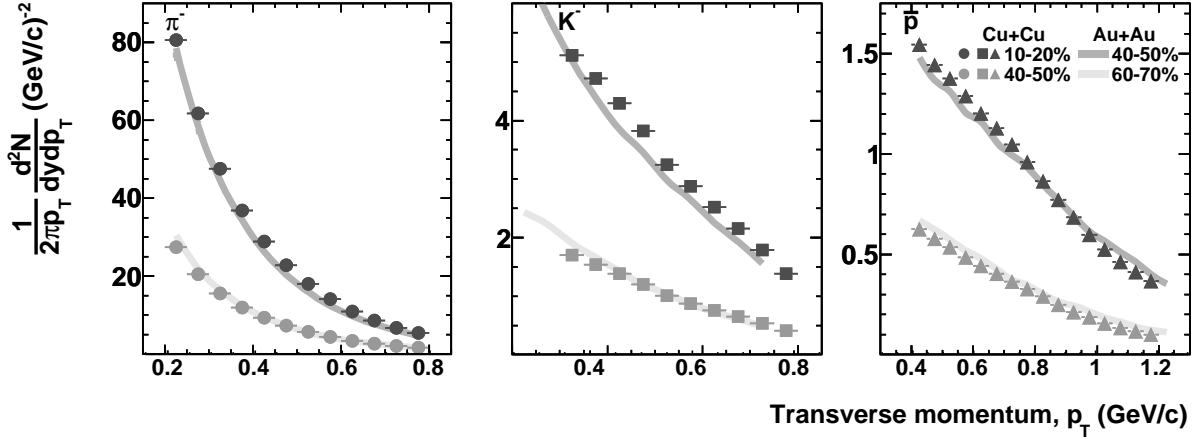}
\caption{\label{fig:CompareSpectra}
Comparison of the spectral shape between Cu+Cu and Au+Au data at
\snn=200~GeV.  Centrality classes are chosen with a similar average
charged hadron multiplicity at mid-rapidity.  Pion (left), kaon
(center) and anti-proton (right) spectra are shown for 10-20\% central
(40-50\%) Cu+Cu (symbols) compared to 40-50\% mid-peripheral (60-70\%)
Au+Au (lines).}
\end{figure*}

The Blast-wave fit results for the temperature of freeze-out are shown in
Fig.~\ref{fig:KineticFO}.  $T_{\rm kin}$ and $\beta$ show similar
dependences as a function of \nch~in both Cu+Cu and Au+Au collisions,
evolving smoothly from the lowest to the highest multiplicity, from \pp~to
central Au+Au.  $T_{\rm kin}$ decreases smoothly with centrality implying
that freeze-out occurs at a lower temperature in more central collisions.  
The similarity of kinetic freeze-out parameters in the events with similar
multiplicity from different colliding species is confirmed by the data alone.
As noted earlier, the particle mean-\pT~increases with increasing \nch,
which is consistent with an increase of radial flow with centrality. 
We note, however, that other physics mechanisms, for example, hard and
semi-hard scatterings, can contribute to higher mean-\pT~values observed
for kaon and proton spectra~\cite{cite:Trainor}.  Direct spectral shape
comparisons of Cu+Cu and Au+Au events from similar multiplicity bins, shown
in Fig.~\ref{fig:CompareSpectra}, show the same \pT-dependencies between
pion spectra from the two systems. The same is seen to hold for the
respective kaon and proton spectra. The middle panel of Fig.~\ref{fig:KineticFO}
shows in addition the chemical 
freeze-out temperatures for different
colliding systems at different energies. Both the chemical freeze-out
and the kinetic freeze-out temperature show similar scaling features,
reflecting the common trends in mean-\pT~and the ratios of $p/\pi$ and
${\rm K}/\pi$, discussed earlier. Similarly, on the left panel of
Fig.~\ref{fig:KineticFO} we observe a
common \nch-dependence for the average radial flow velocity at kinetic
freeze-out. 

A more important observation is that the obtained kinetic freeze-out
parameters for pions, kaons and (anti)protons follow the same trends
with \nch, independent of collision energy, even though the production
cross-sections of the underlying spectra are different.  This observation
suggests that the kinetic freeze-out properties are determined by the
initial state.  Furthermore, a model-dependent connection between the
numbers of produced charged particles and the initial gluon density of
the colliding system~\cite{cite:CGC} can be used to deduce that the
freeze-out properties are most probably determined at the initial
stages of the collision and are driven by the initial energy density.

\subsection{Chemical Properties}

\begin{figure}[!ht]
\includegraphics[width=0.475\textwidth]{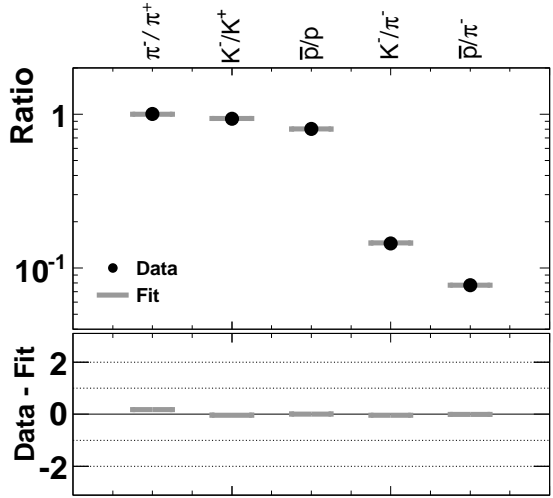}
\caption{\label{fig:ChemicalFit200}
The upper panel shows statistical model fit predictions (grey lines)
for the measured particle ratios (circles) from central 200~GeV Cu+Cu
collisions.  The lower panel illustrates the fit quality by showing
the difference between the measured data and the model prediction in
terms of the number of standard deviations ($N_{\sigma}$) determined
by systematic (data) uncertainty.}
\end{figure}

\begin{figure}[ht]
\vspace{0.5cm}
\includegraphics[width=0.475\textwidth]{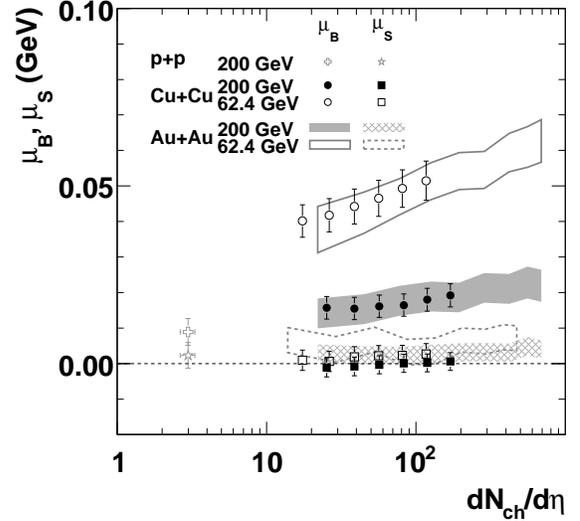}
\caption{\label{fig:ChemicalFO2} Baryon and strangeness chemical potentials, $\mu_{\rm B}$ and $\mu_{\rm S}$,
 as a function of \nch~for 200
and 62.4~GeV in Cu+Cu (symbols) and Au+Au collisions (bands). }
\end{figure}

\begin{figure*}[t]
\includegraphics[width=0.95\textwidth]{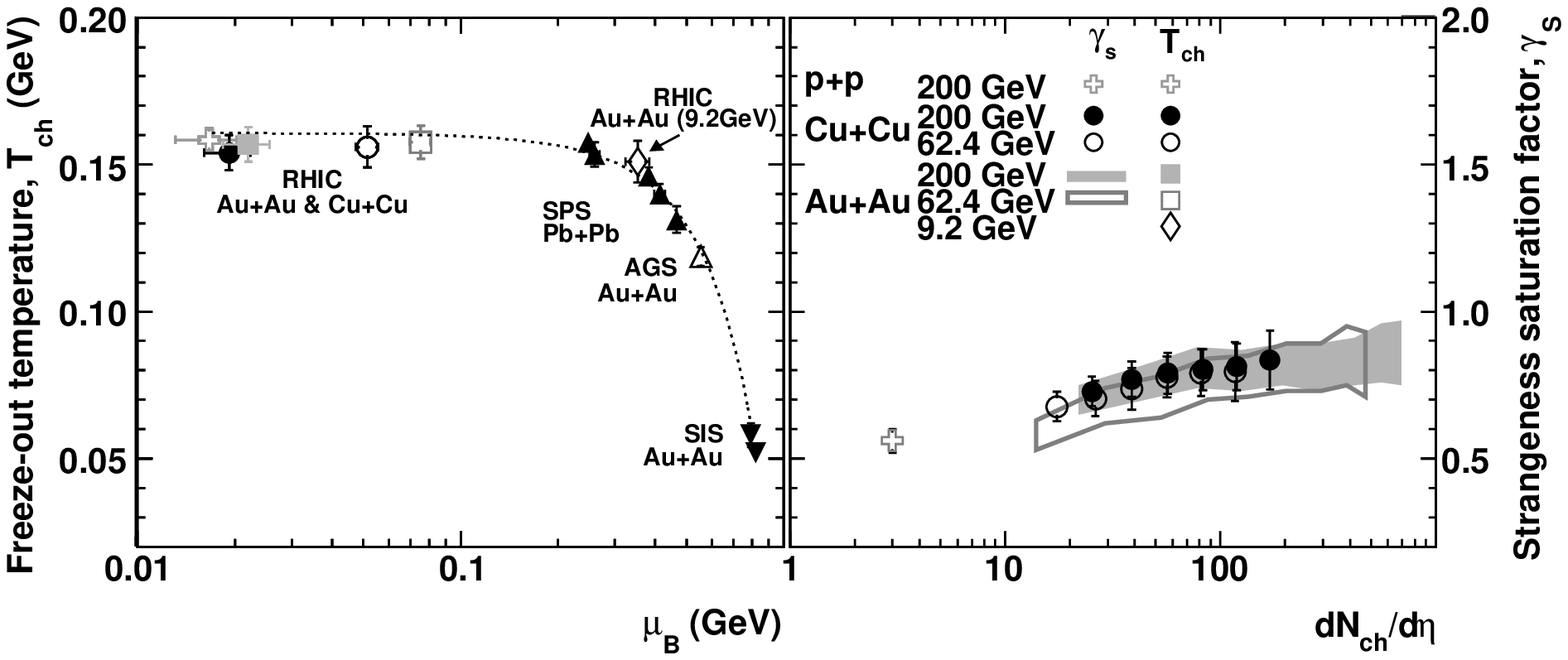}
\caption{\label{fig:ChemicalFO} Left: chemical freeze-out temperature,
$T_{\rm ch}$, as function of the baryon-chemical potential, $\mu_{\rm B}$,
derived for central Au+Au (0-5\% for 200 and 62.4 GeV~\cite{cite:LongPRC} 
and 0-10\% for 9.2 GeV~\cite{cite:STAR9GeV}) and Cu+Cu (0-10\%) collisions.
For comparison, results for minimum bias \pp~collisions at 200~GeV are also
shown along with additional heavy-ion data points compiled for lower
collision energies~\cite{cite:Claymans}.  The dashed line represents
a common fit to all
available heavy-ion data described in the text. Right: strangeness
suppression factor, $\gamma_{\rm S}$, as a function of \nch~for 200
and 62.4~GeV in Cu+Cu (symbols) and Au+Au collisions (bands). }
\end{figure*}

Chemical freeze-out occurs at the stage of the collision when all
inelastic interactions cease and the produced particle composition in
terms of yields is fixed.  Valuable information for this collision
stage can be obtained directly from the experimental results by
forming particle ratios and comparing them across different collision
systems and energies.

The ratios of the different particle yields in Cu+Cu collisions
are further analyzed within the framework of the statistical
model~\cite{cite:StatModel}.  This model describes the chemical
freeze-out of the colliding system by  several fit
parameters: the {\em temperature} at which freeze-out occurs
($T_{\rm ch}$), the {\em cost} of producing matter in terms of
baryon and strangeness chemical potentials ($\mu_{\rm B}$,
$\mu_{\rm S}$), and an additional {\it ad-hoc} parameter, known as
the strangeness suppression  factor, ($\gamma_{\rm s}$),  to reconcile the
lower yield of strange hadrons in collisions
involving smaller species (for example \pp~and d+Au).

These statistical fits are performed on the relative particle
abundances from $\pi^{\pm}$, K$^{\pm}$ and $p$($\overline{p}$) alone.
Figure~\ref{fig:ChemicalFit200} shows an example of the resultant fit
to the identified hadron  ratios from  central \snn=200~GeV Cu+Cu
collisions.  The lower panel of this figure illustrates the fit
quality. We note, that the successful description of the ratios by the
model could not prove the attainment of chemical equilibrium, but
suggests the statistical nature of particle  production in these
collisions~\cite{cite:prc126}.  The results obtained for the
freeze-out parameters are shown in
Figs.~\ref{fig:KineticFO}, \ref{fig:ChemicalFO2}~and~\ref{fig:ChemicalFO}.

Statistical model fits to a wider variety of hadron yields were also
attempted using preliminary results for the  $\Lambda$, $\Xi$ and
$\phi$ particles and anti-particles from 200~GeV Cu+Cu data
from~\cite{cite:AntStrange}.  Including more particles into the model
fits reduces the systematic uncertainty on the extracted parameters
and resulted in parameter values consistent with those  obtained from
fits using $\pi^{\pm}$, K$^{\pm}$ and $p$($\overline{p}$) alone
reported here.  In general, the observed systematic trends in the
freeze-out parameters as a function of the collision centrality are
preserved ~\cite{cite:LongPRC, cite:STARWitePaper}.

Figure~\ref{fig:ChemicalFO} (left panel) shows the evolution of the chemical
freeze-out temperature versus baryon chemical potential in central
heavy-ion collisions from the very low energy SIS data through AGS
and SPS to RHIC (STAR data points only).  The overall evolution of
$T_{\rm ch}$ can be reproduced by the phenomenological model
fit~\cite{cite:Claymans} applied here to all the data points shown
(dashed line).  As the collision energy increases,
 the temperature at freeze-out is found to increase up to
 SPS energies.  This
is followed by a plateau at RHIC energies at a value close to that
of the hadronization temperature expected from lattice QCD
calculations. At RHIC, for all systems and center-of-mass energies,
$T_{\rm ch}$ appears to be universal, as shown in
Fig.~\ref{fig:KineticFO} (middle panel).

The value of the baryon-chemical potential at a given center-of-mass
energy is found to be slightly higher for the larger system, with Au+Au
and Cu+Cu measurements showing common trends with charged hadron
multiplicity (Fig.~\ref{fig:ChemicalFO2}).  We note that, presented in
the same figure, values of strangeness chemical potential are close to
zero with no obvious systematic trends for all energies and colliding
systems studied at RHIC.
Within a given system, $\mu_{\rm B}$ reflects the decrease in net-baryon
density with increasing collision energy from \snn=62.4 to 200~GeV.
This behavior can be observed directly from the particle ratios,
where $\overline{p}/p$ increases as a function of energy
(Fig.~\ref{fig:Ratios}).  For the most central Cu+Cu events we measure
$\overline{p}/p = 0.80\pm0.04$ at 200~GeV, and $0.55\pm0.03$ at 62.4~GeV.
The lack of centrality dependence in the baryon to meson ratios in
Cu+Cu and Au+Au data, points to similar freeze-out temperatures for
the studied systems.  The constant values of $T_{\rm ch}$ at RHIC
energies for collisions with different initial conditions, energy and
net-baryon density, points to a common hadronization temperature of
the systems. 

Another parameter extracted from the fit, which is related to
strangeness production, is the strangeness suppression factor,
$\gamma_{\rm s}$, shown versus \nch~ in Fig.~\ref{fig:ChemicalFO}.
The suppression of strange hadron yields is observed
in smaller systems, such as \pp~and peripheral collisions. Within
statistical models this can be explained by a reduced production
volume~\cite{cite:StrangenessSupp}.  At low beam energies, where
equilibration of $s$ quarks with respect to $u$ and $d$ is not
expected, the suppression is also seen.  We find that within the
systematic errors on the fit parameters the strangeness suppression
factor in Cu+Cu is consistent with that for  Au+Au  for the same number
of charged particles, \nch.  As only charged kaon yields were included
in the fit, this observation is directly related to an absence of any
additional enhancement in ${\rm K}/\pi$ at the same \nch~in the smaller Cu+Cu 
system with respect to the larger Au+Au system as discussed previously.
 
The $\gamma_{\rm s}$ parameter shows a similar increase with centrality
for both systems and energies.  The value of $\gamma_{\rm s}$ approaching
unity for the central Au+Au collisions in the context of thermal model
would imply that the produced strangeness is close to equilibrium.

\section{Summary}
We have presented measurements of identified charged hadron spectra
in Cu+Cu collisions for two center-of-mass energies, 200 and 62.4~GeV.
These new results of $\pi^{\pm}$, K$^{\pm}$ and $p$($\overline{p}$)
have further enriched the variety of low-\pT~spectra at RHIC.  The
data have been studied within the statistical hadronization and
Blast-wave model frameworks in order to characterize the properties
of the final hadronic state of the colliding system as a function of
system size, collision energy and centrality.

These multidimensional systematic studies reveal remarkable
similarities between the different colliding systems.  No additional
enhancement of kaon yields with respect to pions is observed for the
smaller Cu+Cu system compared to Au+Au.  The obtained particle ratios,
mean-\pT~and the freeze-out parameters, including the strangeness
suppression factor, $\gamma_{\rm s}$, are found to exhibit a smooth
evolution with \nch, and similar properties at the same number of
produced charged hadrons are observed for all collision systems and
center-of-mass energies.  The bulk properties studied 
have a strong correspondence with the total particle yield. Within
thermal models this reflects a relation between the energy per
particle at freeze-out and the entropy derived from particle yields,
which reflects the initial state properties for adiabatic expansion.
The baryon chemical potential could in addition be influenced by
the initial valence quark distribution and by baryon transport
during expansion, leading to a more complicated dependence.  The
scaling features of freeze-out properties are not presented at the
same \npart~for lighter and heavier ions as scaling is badly
broken when data measured at different energies are compared. This
suggests that \npart~does not reflect the initial state of the system
accurately.

We thank the RHIC Operations Group and RCF at BNL, the NERSC Center at LBNL and the Open Science Grid consortium for providing resources and support. This work was supported in part by the Offices of NP and HEP within the U.S. DOE Office of Science, the U.S. NSF, the Sloan Foundation, the DFG cluster of excellence `Origin and Structure of the Universe' of Germany, CNRS/IN2P3, STFC and EPSRC of the United Kingdom, FAPESP CNPq of Brazil, Ministry of Ed. and Sci. of the Russian Federation, NNSFC, CAS, MoST, and MoE of China, GA and MSMT of the Czech Republic, FOM and NWO of the Netherlands, DAE, DST, and CSIR of India, Polish Ministry of Sci. and Higher Ed., Korea Research Foundation, Ministry of Sci., Ed. and Sports of the Rep. Of Croatia, Russian Ministry of Sci. and Tech, and RosAtom of Russia.

\end{document}